\documentclass[prl,twocolumn,showpacs,floatfix]{revtex4}

\usepackage{amsmath}
\usepackage{amssymb}

\usepackage{graphicx}
\usepackage{dcolumn}
\usepackage{bm}

\begin{document}

\title{Ground-state of two-dimensional finite electron systems
\\in the Quantum Hall regime}

\author{Henri Saarikoski}
\email[Electronic address:\;]{hri@fyslab.hut.fi}
\author{Ari Harju}

\affiliation{Laboratory of Physics, Helsinki University of Technology,
P.O. Box 1100, FIN-02015 HUT, Finland}

\begin{abstract}
We study electronic structures of quasi-two-dimensional finite electron systems
in high magnetic fields.
The solutions in the fractional quantum Hall regime are interpreted
as quantum liquids of electrons and vortices.
The ground states are classified according to the number of vortices inside
the electron droplet. The theory predicts observable effects due
to vortex formation in the chemical potentials and magnetization of
electron droplets. We compare the transitions in the theory to those
found in electron transport experiments on a quantum dot device
and find significant correspondence.
\end{abstract}

\date{\today}

\maketitle

Studies of finite two-dimensional electron systems in high magnetic fields
often get their inspiration from the remarkable physics of the quantum Hall
effect in two-dimensional electron gas~\cite{fqhe}.
Considerable experimental and theoretical work
has been carried out in particular on quantum dots,
two-dimensional droplets of electrons
in the interface region of semiconductor
heterostructures~\cite{jacak,kouwenhoven2,reimannreview,ashoori}.
Electron transport measurements of quantum dots
have revealed a rich variety of transitions associated with charge
redistribution within the electron droplet in
high magnetic fields~\cite{oosterkamp}.
This has lead to development and study of theoretical models
which could account for the microscopic origin of
these phenomena~\cite{reimann99,yang,oaknin,maksym00}.
Recently a proposal has been put forth that the observed phenomena may 
be caused by emergence of a special state of electronic matter,
a quantum liquid of strongly correlated electrons and
vortices~\cite{saarikoski,toreblad,tavernier,stability}.
In the present work we use this approach to classify the ground states
of the two-dimensional electron droplets in quantum dots and compare
transitions to those found in electron transport measurements.
Our results give also general insight into the internal structure of the
many-body wave function of finite two-dimensional electron systems.

Vortices, rotational flow of currents or matter with a characteristic
cavity at the center, can be found in various natural phenomena where
particles have been set to rotate around a common axis.
Descriptions of vortices exist since antiquity~\cite{odyssey}.
In two-dimensional quantum systems a vortex can be defined in analogue
with classical vortices as a zero in the wave function associated with
a phase change of integer multiple of $2\pi$ for each path enclosing this zero.
In quantum dots the rotation of electrons is induced by external magnetic
field and vortices may form if this rotation is strong.
Vortices are caused by quantization of the magnetic flux through
the electron droplet and they give rise to rotating currents of charge
around density zeros inside the electron droplet.
Vortices create charge deficiency inside the electron droplet
which manifests itself as an increase of the dot area~\cite{reimann99,yang}.
This theory has been based on electronic structure studies
of quantum dots~\cite{saarikoski,tavernier,stability}
as well as on theoretical analogies with
bosonic systems of rotating Bose-Einstein condensates
\cite{toreblad}.
Vortices in electron droplets
are not necessarily bound to electrons as
approximated by the Laughlin wave functions.
The analysis of the internal structure of the many-body
wave function suggests introduction of a more general framework
of interacting system of electrons and off-electron vortex quasi-particles,
where vortex formation is driven by interactions
\cite{stability,dualitymanninen}.

We use an effective-mass approximation in the $xy$-plane
to model the physics of quasi-two-dimensional system of trapped electrons
in vertical quantum dot devices.
The electron-electron
interaction is approximated with Coulomb potential. The Hamiltonian is then
\begin{equation}
H=\sum^N_{i=1}\left(
 \frac{(-i\hbar \nabla_i+e {\bf A} )^2}{2 m^*}
+V_{{\rm c}}(r_i)\right) + \frac{e^2}{4\pi \epsilon} \sum_{i<j}
\frac{1}{r_{ij}} \ ,
\label{hamiltonian}
\end{equation}
where $N$ is the number of electrons,
${V_{{\rm c}}}$ is the external confining potential,
$m^*$ is the effective mass of electrons moving in semiconductor medium,
$\epsilon$ is the dielectric constant, and
${\bf A}$ is the vector potential of the homogeneous magnetic field
which is oriented perpendicular to the $xy$-plane.
In the subsequent discussion the external potential is chosen to be
parabolic $V_c(r)=\frac 12 m^* \omega_0^2r^2$.

In this Letter we use both the mean-field 
spin-density-functional theory (SDFT) and variational quantum Monte Carlo
(VMC) to calculate the electronic structure in the quantum Hall regime.
We use the SDFT in conjunction with local spin density approximation (LSDA)
with a smooth correlation functional~\cite{attaccalite}.
For details of the implementation we refer to
Refs.~\cite{saarikoski,ariproceedings}.
In high magnetic fields the electron droplet is spin-polarized and a stable
stable structure called the maximum density droplet
(MDD) forms~\cite{macdonald}.
It is a finite size precursor of the integer $\nu=1$
quantum Hall state.
Evidence for the MDD formation has been reported in the
experiments~\cite{ashoori,oosterkamp}.
When the magnetic field is increased in the measurements the MDD
breaks down into a lower density droplet in the fractional quantum Hall
(FQH) regime.

Theory predicts that in parabolically confined quantum dots the ground states
in the beyond-MDD (FQH) regime occur only at certain ``magic'' angular momentum
values~\cite{seki96JPSJ}.
As a result the angular momentum as a function of external magnetic field
shows a characteristic staircase structure.
In the MDD breakdown the angular momentum
increase with respect to the MDD state $\Delta L=L-L({\rm MDD})$
strongly depends on the number of electrons
in the system~\cite{ariproceedings}.
For $N\le 12$ the electron in the center
is moved to the outer edge giving $\Delta L=N$ and a vortex hole emerges
at the center of the dot.
For $N>12$ a vortex emerges at a finite distance from the center~\cite{yang}.
This change in the breakdown mechanism means that
vortices in large electron systems tend not to localize.
However, if the symmetry of the external potential is broken localized
vortices may form in the particle and current densities~\cite{stability}.
High angular momentum states correspond to multi-vortex configurations
in the FQH regime. Since they are beyond reach for
exact diagonalization techniques for $N > 10$
we use the SDFT to analyse the electronic structure of these states.

\begin{figure}
\includegraphics[width=.95\columnwidth]{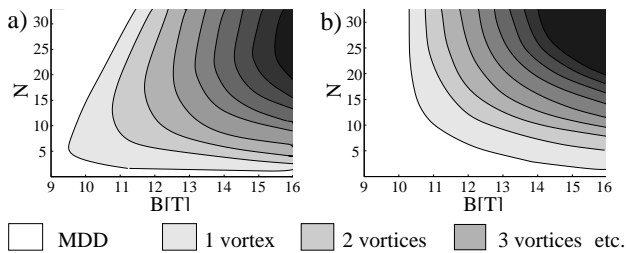}
\caption{Mean-field SDFT phase diagrams of the ground state of parabolically
confined electron droplets in high magnetic fields. The ground states
are classified according to the number of vortices inside the electron
droplet (gray-scale).
The confining potential is $\hbar\omega_0= 5\;{\rm meV}$ in a) and
$7.67 N^{-1/7}$ meV in b).
In the upper right hand corner of the right diagram the number of vortices
is greater than 8.
}
\label{fig:phasediagram}
\end{figure}

In the experimental realizations of quantum dots
the area of the dot has been found to increase with the
gate voltage suggesting that the electron density in the
dot remains constant~\cite{austing}.
In zero magnetic field this implies a confining potential
that scales as $\hbar \omega_0\sim N^{-1/4}$~\cite{koskinen97}.
However, the magnetic field exerts additional squeezing effect on the electrons
which is counteracted by interactions.
We have found that approximately
constant electron density in the calculations is obtained
by a potential scaling $\hbar \omega_0\sim N^{-1/7}$.
Fig.~\ref{fig:phasediagram} shows phase diagrams of the ground states
in the SDFT.
The ground states are classified according to number of vortices inside
the electron droplet by using a conditional single-determinant
wave functions constructed from Kohn-Sham orbitals~\cite{saarikoski2}.

For qualitative understanding of the structure of the phase diagram
we interpret the solutions in the FQH domain as a quantum liquid of electrons
and hole-like vortex quasi-particles.
The MDD state assigns one Pauli vortex at each electron
position. As the MDD reconstructs the number of vortices in
the system increases by one and subsequent transitions involve
emergence of more off-electron vortices.
Since vortices carry magnetic flux quanta $\Phi_0$
the MDD state breaks down approximatively when the magnetic flux $\Phi=BA$
through the MDD of area $A$ exceeds $(N+1)\Phi_0$ and
an additional off-electron vortex emerges in the electron droplet.
The compactness of the MDD state and the observed constancy of the electron
density $n$ with respect to $N$ gives $N=nA$ and 
$B=(1+1/N)\Phi_0 n$ for the MDD reconstruction.
The upper boundary of the MDD is therefore approximatively
constant for high $N$ in accord with the experiments~\cite{oosterkamp}.
In the addition of one vortex in the MDD
the relative increase of the dot area is approximately $1/N$.
This increase of the dot area via vortex formation counteracts 
the squeezing effect of the magnetic field.
Assuming now constant $n$ for the beyond-MDD states
the change in the $B$ required for addition of subsequent off-electron vortices
in the droplet is approximately $\Delta B=\Phi_0n/N$.
The calculations show that widths of the different vortex phases are
to a good accuracy inversely proportional to $N$
which gives credence to this simplified picture
(see Fig.~\ref{fig:phasediagram}b).
The above reasoning implies also that constancy
of the electron density in the droplet leads to narrowing of the MDD window for
high electron numbers. The lower boundary of the MDD is determined either
from polarization of the electron droplet or flipping of one vortex
from parallel to anti-parallel orientation with respect to the magnetic field.
The latter condition involves a constant flux change of $2\Phi_0$
with respect to the MDD state.
This leads to narrowing of the MDD window because the area of the droplet
increases with $N$ in a totally polarized system.
However, for realistic Zeeman-coupling strengths
the electron droplet loses total spin-polarization in the low-field limit
before antivortices emerge.
This shrinks the MDD window further.

\begin{figure}
\includegraphics[width=.90\columnwidth]{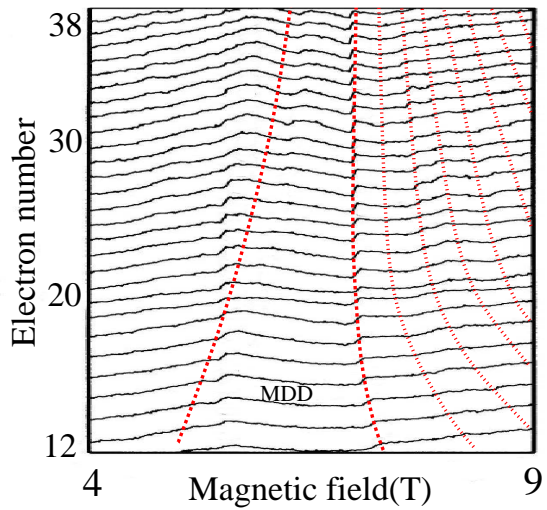}
\caption{
Current peaks in the electron transport experiments and transitions
in the SDFT (red lines). The dashed lines denote the MDD boundaries.
The right dashed line and the dotted lines correspond to transitions
associated with emergence of off-electron vortices one-by-one
inside the electron droplet. The experimental data is from Fig.~2b
in Ref.~\cite{oosterkamp}.
}
\label{comparison}
\end{figure}

\begin{figure}
\includegraphics[width=.98\columnwidth]{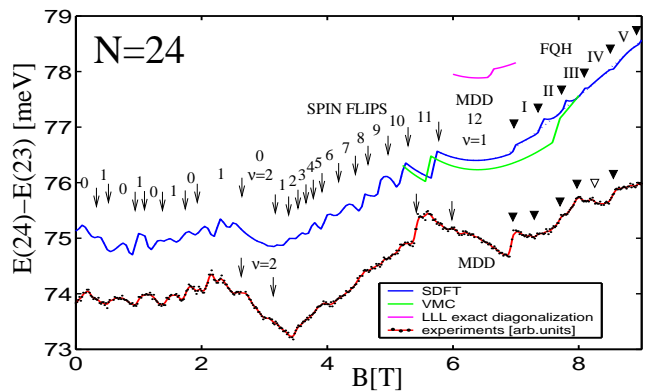}
\caption{
Chemical potential of the 24-electron quantum dot calculated with
the SDFT and compared to the experiments
from Ref.~\cite{oosterkamp} and
VMC and LLL exact diagonalization results in the vicinity of the MDD window.
Noise in the experimental data has been reduced by using a
Gaussian filter.
The numbers between the arrows indicate the total spin.
The roman numbers between the triangles indicate the number of vortices
inside the electron droplet in the fractional quantum Hall regime.
Charge-density-wave solutions in the SDFT (dotted line) have been
discarded as unphysical.
}
\label{chempot}
\end{figure}

We now compare these theoretical results with experimental data from
electron transport measurements.
Oosterkamp and coworkers measured electron transport
through a vertical quantum dot device in the quantum Hall regime
\cite{oosterkamp}. Up to this date these experiments
give the best available electron transport data for a single-dot device.
Fig.~\ref{comparison} shows
chemical potentials in these experiments for electron numbers
$N=12$ to $39$,
the MDD window boundaries in the SDFT as well as the transitions
associated with emergence of off-electron vortices.
The gate voltage dependence of the external confinement
in the quantum dot device is taken into account by using
confining potential $\hbar\omega_0=5.70N^{-1/7}\rm{meV}$ in the SDFT.
For material parameters we use $m^*=0.067$, $\epsilon=12.4$,
and effective gyromagnetic ratio $g^*=-0.44$.
Despite the simple form of the inter-electron potential in our theoretical
model the transitions in the beyond-MDD domain fit well into the experimental
data. This can be understood from the fact that the transitions are induced
by the magnetic flux quantization through the relatively compact
electron droplet.
In Fig.~\ref{chempot} the chemical potential of
the 24-electron quantum dot is calculated with the SDFT, the VMC,
and the exact diagonalization in the lowest Landau level (LLL).
The results are compared to the experimental data in Ref.~\cite{oosterkamp}.
The correspondence between the theory and experiments
is good and different Quantum Hall regimes can be identified by
comparing the two sets of data.
We find also that the theoretical results are consistent with the
observed narrowing of the MDD window from about 1T at $N=20$ to
0.5 T at $N=39$ (see Fig.~\ref{comparison}).
In the SDFT the partially polarized states
in the spin-flip region
have a MDD-like compact structure for both orientations of the $z$-component
of the spin. Therefore 
the states before the MDD around $6\; {\rm T}$ for high $N$
in Fig.~\ref{comparison}
could be partially polarized states and the MDD window may be
smaller than that identified in Ref.~\cite{oosterkamp}.
In the theory the increase in the dot area at the MDD breakdown is
approximately $1/N$=3.3\% at $N=30$.
This can be contrasted to around 10\% 
reported in the experiments~\cite{oosterkamp}.
However, uncertainty in the experimental result is high and
there exists no analysis of a possible $N$ dependence~\cite{guyprivate}.

The electron transport data shows features which may be due to
correlation effects beyond our mean-field SDFT.  These may include the
transition associated with the open triangle in Fig. \ref{chempot} and
fluctuations in the data between the second and third triangle. In
addition, the LLL theory of Ref.~\cite{oaknin} suggest that for small
particle numbers ($\le 100$), the first ground state after MDD would
have partial spin polarization. However, higher Landau levels might
have effect on this \cite{sami}. In actual quantum dot realizations,
effects due to finite thickness of the electron gas, screening of the
interaction potential, image charges, and non-parabolic terms in the
external potential may also cause deviations from our results
\cite{nishi}.

\begin{figure}
\includegraphics[width=.9\columnwidth]{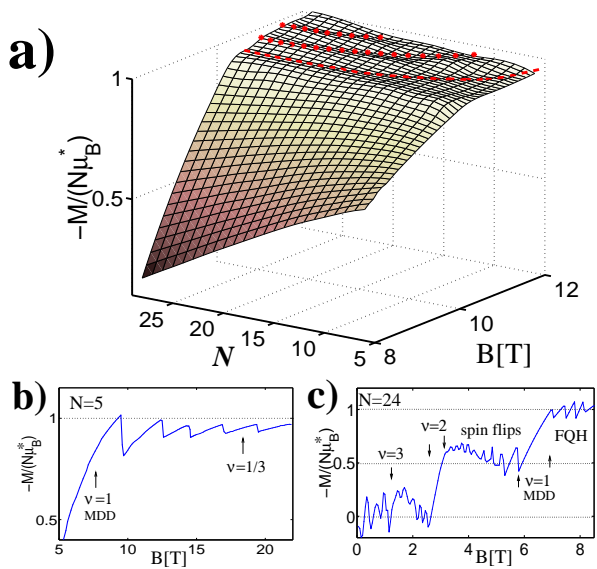}
\caption{a)
Ground state magnetization per electron calculated with the SDFT.
The confining potential is $7.67 N^{-1/7}{\rm meV}$.
The beyond-MDD states are characterized by a plateau region
in this plot, where magnetization per electron is close to $\mu_B^*$.
The dashed and the dotted lines correspond to the same transitions as in
Fig.~\ref{comparison}.
b) and c) Detail plots of the oscillations in the magnetization of
the 5-electron and 24-electron droplets, respectively.
The finite-size precursor of the
$\nu=1/3$ quantum Hall state is identified by using
conditional wave functions~\cite{saarikoski}.
The confinement strength is 5 meV in a)
and 3.62 meV in b).
The oscillations in the plot a) are smoothed out due to 
use of finite temperature.
}
\label{magnetizationfig}
\end{figure}

In the electron transport experiments
the slope of the chemical potential is the magnetization
difference $M(N,B)-M(N-1,B)$, where $M=-\partial E/\partial B$.
Since the magnetic field couples to the angular momentum $L$, there is a
jump in the magnetization at transitions associated with change in $L$.
In the MDD state the angular momentum increases with $N$ as $N\to N+1$.
Therefore the slopes of the chemical potentials in the MDD domain show
typical decreasing pattern as $N$ increases at fixed $B$.
SDFT calculations indicate that magnetization per electron in the
FQH regime is close to the effective Bohr magneton
$\mu_B^*=m_e/m^* \mu_B=0.864\;{\rm meV/T}$ (see Fig.~\ref{magnetizationfig}).
This is a consequence of emergence of additional vortex
quasi-particles in the electron droplet as the magnetic field increases.
This behaviour gives rise to positive slopes 
in the chemical potentials in the beyond-MDD domain
which is in accord with experimental data
(see {\em e.g.} slopes of the chemical potentials at $B=9$ T in
Fig.~\ref{comparison}).
Detail plots of the magnetization for $N=5$ and $N=24$
in the mean-field theory
(Figs.~\ref{magnetizationfig}b and~\ref{magnetizationfig}c)
reveal also oscillations
in the droplet magnetization as the number of vortices increases one-by-one
\cite{stability}. The sawtooth-like oscillations in lower magnetic fields
in Fig.~\ref{magnetizationfig}c are manifestations of the de Haas--van Alphen
effect.
The overall behaviour is similar to that found in 
direct measurements of magnetization of dot mesas
~\cite{schwarz}. This method could also provide a way to detect
oscillations in the FQH regime.

To conclude, we have performed theoretical calculations for
finite electron droplets in the Quantum Hall regime.
Our model theory predicts emergence of a quantum liquid of electrons and
off-electron vortices in high magnetic fields. The pattern of transitions
found within the theory is consistent with experimental data
from electron transport measurements.
However, for greater qualitative understanding of the phenomena
we call for more accurate electron transport
experiments in the quantum Hall regime, more accurate computational methods,
and more realistic modeling of quantum dot systems in the case of
many-electron systems.
It has been suggested that addition of cusps to smooth LSDA functionals
incorporates more FQH correlations \cite{lublin}.
Another approach towards testing of the validity of theoretical predictions
could be a direct visualization of the electron density in a
quantum dot. Recent developments with scanned probe imaging techniques
\cite{westervelt,fallahi} could yield a way to image
localized vortices in electron droplets.

We are grateful to authors of Ref. \cite{oosterkamp}
for permission to reproduce some of the published electron transport data.
We thank E. R\"as\"anen, S.~M. Reimann, L. Kouwenhoven, D. G. Austing, 
Y. Nishi, M. Marlo-Helle, G.~S. Jeon, and M. Puska for fruitful discussions
and K. Saloriutta for
computational work in producing Fig.~\ref{magnetizationfig}b.
This work has been supported by Academy of Finland through the Centre
of Excellence Program (2000-2005).


\begin{thebibliography}{11}
\bibitem{odyssey}
Charybdis is described as a giant whirlpool in Homer's epic poem
the Odyssey, book XII.
\end{thebibliography}
\end{document}